\documentclass[sigplan,nonacm,screen]{acmart}

\usepackage{xcolor}
\usepackage{xspace}
\usepackage{algorithm}
\usepackage[noend]{algpseudocode}
\usepackage{wrapfig}
\usepackage{subcaption}
\usepackage[capitalize]{cleveref}

\crefformat{section}{#2\S#1#3}
\crefname{figure}{Fig.}{Figs.}
\crefname{table}{Tab.}{Tabs.}

\algtext*{EndWhile} 
\algtext*{EndIf} 
\algtext*{EndFor} 

\newcommand{\sys}{\textsc{\textsf{MoE-Lightning}}\xspace}
\newcommand{\pipe}{\textsc{\textsf{CGOPipe}}\xspace}
\newcommand{\cm}{\textsc{\textsf{HRM}}\xspace}

\begin{document}

\title{\sys: High-Throughput MoE Inference on Memory-constrained GPUs}

\author{Shiyi Cao}
\email{shicao@berkeley.edu}
\affiliation{%
  \institution{UC Berkeley}
  \city{Berkeley}
  \state{CA}
  \country{USA}
}
\author{Shu Liu}
\email{lshu@berkeley.edu}
\affiliation{%
  \institution{UC Berkeley}
  \city{Berkeley}
  \state{CA}
  \country{USA}
}
\author{Tyler Griggs}
\email{tgriggs@berkeley.edu}
\affiliation{%
  \institution{UC Berkeley}
  \city{Berkeley}
  \state{CA}
  \country{USA}
}
\author{Peter Schafhalter}
\email{pschafhalter@berkeley.edu}
\affiliation{%
  \institution{UC Berkeley}
  \city{Berkeley}
  \state{CA}
  \country{USA}
}
\author{Xiaoxuan Liu}
\email{xiaoxuan_liu@berkeley.edu}
\affiliation{%
  \institution{UC Berkeley}
  \city{Berkeley}
  \state{CA}
  \country{USA}
}
\author{Ying Sheng}
\email{ying1123@stanford.edu}
\affiliation{%
  \institution{Stanford}
  \city{Palo Alto}
  \state{CA}
  \country{USA}
}
\author{Joseph E. Gonzalez}
\email{jegonzal@berkeley.edu}
\affiliation{%
  \institution{UC Berkeley}
  \city{Berkeley}
  \state{CA}
  \country{USA}
}
\author{Matei Zaharia}
\email{matei@berkeley.edu}
\affiliation{%
  \institution{UC Berkeley}
  \city{Berkeley}
  \state{CA}
  \country{USA}
}
\author{Ion Stoica}
\email{istoica@berkeley.edu}
\affiliation{%
  \institution{UC Berkeley}
  \city{Berkeley}
  \state{CA}
  \country{USA}
}


\begin{abstract}
Efficient deployment of large language models, particularly Mixture of Experts (MoE) models, on resource-constrained platforms presents significant challenges in terms of computational efficiency and memory utilization. The MoE architecture, renowned for its ability to increase model capacity without a proportional increase in inference cost, greatly reduces the token generation latency compared with dense models. However, the large model size makes MoE models inaccessible to individuals without high-end GPUs.
In this paper, we propose a high-throughput MoE batch inference system, \sys, that significantly outperforms past work. \sys introduces a novel CPU-GPU-I/O pipelining schedule, \pipe, with \emph{paged weights} to achieve high resource utilization, and a performance model, \cm, based on a Hierarchical Roofline Model we introduce to help find policies with higher throughput than existing systems.
\sys can achieve up to \(10.3\times\) higher throughput than state-of-the-art offloading-enabled LLM inference systems for Mixtral 8x7B on a single T4 GPU (16GB). When the theoretical system throughput is bounded by the GPU memory, \sys can reach the throughput upper bound with 2--3\(\times\) less CPU memory, significantly increasing resource utilization. \sys also supports efficient batch inference for much larger MoEs (e.g., Mixtral 8x22B and DBRX) on multiple low-cost GPUs (e.g., 2--4 T4s).
\end{abstract}

\maketitle 
\pagestyle{plain} 

\section{Introduction}

Mixture of Experts (MoE)~\cite{shazeer2017outrageously,dbrx,deepseek,mixtral} is a paradigm shift in the architecture of Large Language Models (LLMs) that leverages sparsely-activated expert sub-networks to enhance model performance without significantly increasing the number of operations required for inference. Unlike dense models~\cite{llama, opt, scao2022bloom}, where all model parameters are activated for each input, MoE models activate only a subset of experts, thereby improving computational efficiency.

While the MoE models achieve strong performance in many tasks~\cite{deepseek, mixtral}, unfortunately, their deployment is challenging due to the significantly increased memory demand for the same number of active parameters. For example, the Mixtral 8x22B model~\cite{mixtral22b} requires over 256 GB of memory for the parameters of the expert feed-forward network (FFN), which is \(4-5\times\) higher than the memory requirements of dense models that require similar FLOPs for inference.

In this paper, we study how to achieve \emph{high-throughput} MoE inference with limited GPU memory. We are focusing on off-line, batch-processing workloads such as model evaluation~\cite{liang2022holistic}, synthetic data generation~\cite{dubey2024llama}, data wrangling~\cite{narayan2022can}, form processing~\cite{chen2021spreadsheetcoder}, and LLM for relational analytics~\cite{liu2024optimizing} where higher inference throughput translates into lower total completion time.

The common approach for memory-constrained batch inference is to offload model weights~\cite{aminabadi2022deepspeed, huggingfaceAccelerate} and key-value tensors of earlier tokens (KV cache)~\cite{sheng2023flexgen} --- which are needed for generating the next token -- to CPU memory or disk. Then, they are loaded layer-by-layer to the GPU for computation.

\begin{figure}[ht]
    \vspace{-1em}
    \centering
    \includegraphics[width=0.93\linewidth]{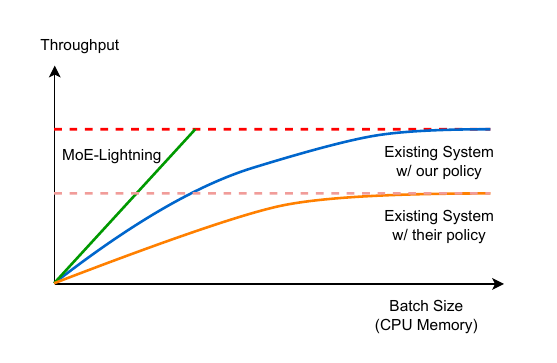}
    \caption{\sys achieves higher throughput with far less CPU memory, enabled by \pipe and \cm.}
    \label{fig:intro}
\end{figure}

Unfortunately, existing solutions fall short of effectively overlapping computations with data transfers between CPU and GPU. 
For instance, the GPU may remain idle as it awaits a small yet crucial piece of data 
such as intermediate results for the upcoming batch. At the same time, transferring the weights for subsequent layers may take a long time and potentially block both the GPU and CPU from processing further tasks, leading to under-utilization of all the resources.

As a result, efficient MoE inference for throughput-oriented workloads using limited GPU memory remains challenging. We find that increasing I/O utilization and other resource utilization is critical in achieving high throughput.
For example, \cref{fig:intro} illustrates the relationship between CPU memory and achievable token generation throughput for different systems with fixed GPU memory (less than the model size) and CPU-to-GPU memory bandwidth. When a layer's weights are loaded onto the GPU, a common strategy to increase throughput is to process as many requests as possible to amortize the I/O overhead of weights' transfer~\cite{sheng2023flexgen}. However, this increases CPU memory usage as additional space is required to store the KV cache for all requests.
Consequently, lower I/O utilization means higher I/O overhead of weights' transfer, requiring greater CPU memory to reach peak generation performance; otherwise, the GPU will be under-utilized as suggested by the blue line in \cref{fig:intro}.

While improving resource utilization is crucial for achieving high-throughput inference with limited GPU memory, achieving this raises several challenges.
\textbf{\textit{First}}, we need to effectively schedule the computation tasks running on CPU and GPU, together with the transfers of various inputs (e.g., experts weights, hidden states, and KV cache),  such that to avoid computation tasks waiting for transfers or the other way around. 
\textbf{\textit{Second}}, as indicated by the orange line in \cref{fig:intro}, the existing solutions~\cite{sheng2023flexgen} tend to generate sub-optimal policies with smaller GPU batch sizes which lead to resource under-utilization. Fundamentally, these solutions fail to take into account that changes in the workload can lead to changes in the bottleneck resource. 

To address these two challenges, we developed a new inference system, \sys, which consists of two new components. The first component is \textbf{\pipe}, a pipeline scheduling strategy that overlaps GPU computation, CPU computation and various I/O events efficiently so that computation is not blocked by I/O events and different I/O events won't block each other. This way, \pipe can significantly improve the system utilization.
The second component is \textbf{Hierarchical Roofline Model (HRM)} which accurately models how different components in an inference system interact and affect application performance under various operational conditions. 

In summary, this paper makes the following contributions:
\begin{itemize}
    \item \pipe, a pipeline scheduling strategy that efficiently schedules various I/O events and overlaps CPU and GPU computation with I/O events. 
    By deploying \emph{weights paging}, \pipe reduces pipeline bubbles, significantly enhancing throughput and I/O efficiency compared with existing systems (\cref{sec:pipeline}).
    \item \cm, a \textit{general} performance model for LLM inference 
    which extends the Roofline Model~\cite{WilliamsWP09}. \cm can easily support different models, hardware, and workloads, and has near-zero overhead in real deployments, 
    without the need for extensive data fitting (might take hours or days) as needed in FlexGen (\cref{sec:performance_model}).
    \item An in-depth performance analysis for MoE models based on our extended HRM which identifies various performance regions where specific resource becomes the bottleneck (\cref{sec:pa}).
\end{itemize}

We evaluate \sys on various recent popular MoE models (e.g., Mixtral 8x7b, Mixtral 8x22B, and DBRX) on different hardware settings (e.g., L4, T4, 2xT4, and 4xT4 GPUs) using three real workloads. When compared to the best of the existing systems, \sys can improve the generation throughput by up to \(10.3\times\) (without request padding) and \(3.5\times\) (with request padding) on a single GPU. When Tensor-parallelism is enabled, \sys demonstrates \textbf{\textit{super-linear scaling}} in generation throughput (\cref{sec:eval}).
\section{Background}
\subsection{Mixture of Experts}\label{sec:moe_bg}
Large Language Models (LLMs) have significantly improved in performance due to the advancements in architecture and scalable training methods.
In particular, Mixture of Experts (MoE) models have shown remarkable improvements in model capacity, training time, and model quality~\cite{shazeer2017outrageously,glam,deepseek,dbrx,mixtral,lepikhin2020gshard}, revitalizing an idea that dates back to the early 1990s~\cite{JacobsJNH91,jordan1994hierarchical} where ensembles of specialized models are used in conjunction with a gating mechanism to dynamically select the appropriate ``expert'' for a given task.

The key idea behind MoE is a gating function that routes inputs to specific experts within a larger neural network. Each expert is specialized in handling particular types of inputs.
The gating function selects only a subset of experts to process an input, which allows LLMs to scale the number of parameters without increasing inference operations.

\begin{figure}[ht]
    \centering
    \includegraphics[width=0.25\textwidth]{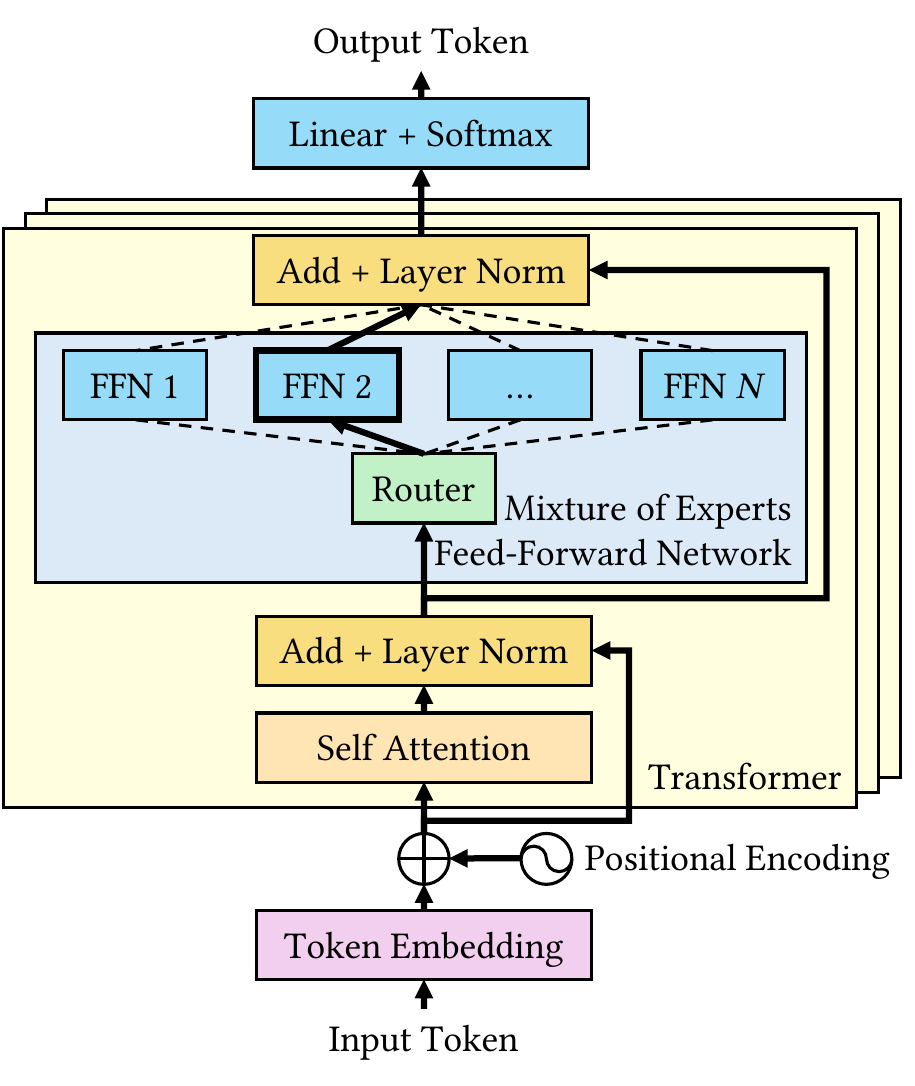}
    \caption{Architecture of a Mixture of Experts in Large Language Models.}
    \label{fig:moe_architecture}
    \vspace{-1em}
\end{figure}

MoE models adopt a conventional LLM architecture, which uses learned embeddings for tokens and stacked transformer layers.
MoE LLMs typically modify the Feed-Forward Network (FFN) within a transformer layer by adding a gating network that selects expert FFNs, usually implemented as multi-layer perceptrons, to process the input token~\cite{glam,zhou2022mixture,chen2023lifelong}.
These designs can surpass traditional dense models~\cite{chatbot-arena, deepseek, mixtral} in effectiveness while being more parameter-efficient and cost-effective during training and inference.

Despite their advantages, the widespread use of MoE models faces challenges due to the difficulties in managing and deploying models with extremely high parameter counts that demand substantial memory. Thus, our work aims to make MoE models more accessible to those lacking extensive high-end GPU resources.

\subsection{LLM Inference}

LLMs are trained to predict the conditional probability distribution for the next token, $P(x_{n+1} \mid x_1, \ldots, x_n)$, given a list of input tokens $(x_1, \ldots, x_n)$. When deployed as a service, the LLM takes in a list of tokens from a user request and generates an output sequence $(x_{n+1}, \ldots, x_{n+T})$. The generation process involves sequentially evaluating the probability and sampling the token at each position for $T$ iterations. The stage where the model generates the first token $x_{n+1}$ given the initial list of tokens $(x_1, \ldots, x_n)$, is defined as the \textit{prefill stage}. In the prefill stage, at each layer, the input hidden states to the attention block will be projected into the query, key, and value vectors. The key and value vectors will be stored in the KV cache.
Following the prefill stage is the \textit{decode stage}, where the model generates the remaining tokens $(x_{n+2}, \ldots, x_{n+T})$ sequentially. When generating token $x_{n+2}$, all the KV cache of the previous tokens $(x_1, \ldots, x_{n+1})$ will be needed, and the token $x_{n+2}$'s key and value at each layer will be appended to the KV cache.

The auto-regressive nature of LLM generation, where tokens are generated sequentially, can lead to sub-optimal device utilization and decreased serving throughput~\cite{pope2023efficiently}. Batching is a critical strategy for improving GPU utilization: \cite{yu2022orca} proposed continuous batching which increases the serving throughput by orders of magnitude. Numerous studies have developed methods to tackle associated challenges such as memory fragmentation~\cite{kwon2023efficient} and the heavy memory pressure imposed by the KV cache~\cite{he2024fastdecode, sheng2023flexgen, juravsky2024hydragen}.
The scenario of limited GPU memory introduces further challenges, especially for large MoE models, as it requires transferring large amounts of data between the GPU and CPU for various computational tasks with distinct characteristics. Naive scheduling of the computation task and data transfer can result in poor resource utilization. This paper explores how each resource in a heterogeneous system affects LLM inference performance and proposes efficient scheduling strategies and system optimizations to enhance resource utilization.
\section{Performance Analysis}\label{sec:pa}
In this section, we introduce a Hierarchical Roofline Model (HRM) (\cref{sec:hrm}) extended from the classical Roofline Model~\cite{WilliamsWP09}, which we use to conduct a theoretical performance analysis for MoE inference (\cref{sec:case_study}). It also serves as the basics of our performance model used in scheduling policy search, which will be discussed in \cref{sec:performance_model}.
The Hierarchical Roofline Model extends the original Roofline Model for multicore architectures~\cite{WilliamsWP09} to provide a stronger model of heterogeneous computing devices and memory bandwidth.
We further identify two additional turning points that define settings where the computation is best done on CPU instead of GPU and where the application is GPU memory-bound or CPU memory-bound, providing explicit explanations for how LLM inference performance will be affected by different resource limits in the system.

\subsection{Roofline Model}
We will start with the original Roofline Model~\cite{WilliamsWP09}, which provides a visual performance model to estimate the performance of a given application by showing inherent hardware limitations and potential opportunities for optimizations.
It correlates a system's peak performance and memory bandwidth with the operational intensity of a given computation, where Operational Intensity ($I$) denotes the ratio of the number of operations in FLOPs performed to the number of bytes accessed from memory, expressed in FLOPs/Bytes.
\begin{wrapfigure}{r}{0.25\textwidth}
    \vspace{-0.5em}
    \hspace{-1.3em}
    \includegraphics[width=0.25\textwidth]{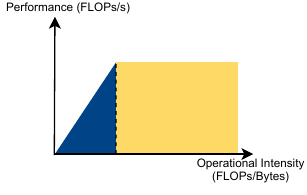}
    \vspace{-1.4em}
\end{wrapfigure}
The fundamental representation in the Roofline Model is a performance graph, where the x-axis represents operational intensity \(I\) in FLOPs/byte and the y-axis represents performance $P$ in FLOPs/sec. The model is graphically depicted by two main components:

\textbf{Memory Roof:} It serves as the upper-performance limit indicated by memory bandwidth. It is determined by the product of the peak memory bandwidth ($B_{\text{peak}}$ in Bytes/sec) and the operational intensity ($I$). Intuitively, if the data needed for the computation is supplied slower than the computation itself, the processor will idly wait for data, making memory bandwidth the primary bottleneck.
The memory-bound region (in blue) of the roofline is then represented by:
\begin{equation}
    P \leq B_{\text{peak}} \times I
\end{equation}
where $P$ is the achievable performance. 

\textbf{Compute Roof:} This represents the maximum performance achievable limited by the machine's peak computational capability ($P_{\text{peak}}$). It is a horizontal line on the graph (top edge of the yellow region), independent of the operational intensity, indicating that when data transfer is not the bottleneck, the maximum achievable performance is determined by the processor's computation capability. The compute-bound part (yellow region) is then defined by:
\begin{equation}
    P \leq P_{\text{peak}}
\end{equation}

The turning point is the intersection of the compute and memory roofs, given by the equation:
\begin{equation}
    \bar{I} = \frac{P_{\text{peak}}}{B_{\text{peak}}} \label{eq:tp1}
\end{equation}
defines the critical operational intensity $\bar{I}$. Applications with $I \geq \bar{I}$ are typically \emph{compute-bound}, while those with $I < \bar{I}$ are \emph{memory-bound}.

In practice, analyzing an application's placement on the roofline model helps identify the critical bottleneck for performance improvements. Recent works~\cite{yuan2024llm} analyze different computations (e.g., softmax and linear projection) in LLM using the Roofline Model.

\subsection{Hierarchical Roofline Model}\label{sec:hrm}
While the original Roofline Model demonstrates great power for application performance analysis, 
it is not enough for analyzing applications such as LLM inference that utilize diverse computing resources (e.g., CPU and GPU) and move data across multiple memory hierarchies (e.g., GPU HBM, CPU DRAM, and Disk storage).

Consider a system with \(n\) levels of memory hierarchies. Each level \(i\) in this hierarchy is coupled with a computing processor. The peak bandwidth at which the processor at level \(i\) can access the memory at the same level is denoted by \(B_{\text{peak}}^{i}\). Additionally, the peak performance of the processor is denoted by \(P_{\text{peak}}^{i}\)\footnote{In this paper we assume when $i<j$, \(P_{\text{peak}}^{i} \geq P_{\text{peak}}^{j}\) and \(B_{\text{peak}}^{i} \geq B_{\text{peak}}^{j}\).}. 

\begin{definition}[General Operational Intensity]
To consider different memory hierarchies, we define the general operational intensity \(I_x^i\) of the computation task \(x\) as the ratio of the number of operations in FLOPs performed by \(x\) to the number of bytes accessed from memory at level \(i\).
\end{definition}

For computation \(x\) executed at level \(i\) in the HRM, we can define its compute and memory roofs similarly as in the original Roofline Model:

\begin{itemize}
    \item \textbf{Compute Roof at level \(i\)}:
    \begin{equation}
    P_x^i \leq P_{\text{peak}}^i \label{eq:cr_i}
    \end{equation}
    This represents the maximum computational capability at level \(i\), independent of the operational intensity.

    \item \textbf{Memory Roof at level \(i\)}:
    \begin{equation}
    P_x^i \leq B_{\text{peak}}^i \times I_x^i \label{eq:mr_i}
    \end{equation}
\end{itemize}

More importantly, in HRM, there is also the memory bandwidth from level $j$ to level $i$, denoted as \(B_{\text{peak}}^{j, i}\), which will define another memory roof for computation $x$ that is executed on level $i$ and transfers data from level $j$:

\begin{itemize}
    \item \textbf{Memory Roof from level \(j\) to \(i\)}:
    \begin{equation}
    P_x^i \leq B_{\text{peak}}^{j, i} \times I_x^j \label{eq:mr_ji}
    \end{equation}
\end{itemize}

Therefore, if computation $x$ is executed on level $i$, data from level $j$ needs to be fetched, and the peak performance will be bounded by the three roofs listed above (\cref{eq:cr_i,eq:mr_i,eq:mr_ji}):
\begin{align}
    P_x^i &= \min(P_{\text{peak}}^i, B_{\text{peak}}^i \times I_x^i, B_{\text{peak}}^{j, i} \times I_x^j) \label{eq:P_i_j}
\end{align}

If operator $x$ is executed on level $i$ without fetching data from other levels, it reduces to the traditional roofline model and can achieve: 
\begin{align}
P_x^i &= \min(P_{\text{peak}}^i, B_{\text{peak}}^i \times I_x^i) \label{eq:P_i}
\end{align}

\paragraph{Turning Points}
Intuitively, our HRM introduces more memory roofs that consider cross-level memory bandwidth and compute roofs for diverse processors. This results in more ``turning points'' than in the original Roofline Model, which define various performance regions where different resources are the bottleneck. Analyzing these turning points is crucial for understanding the performance upper bound of an application under different hardware setups and computational characteristics.

For example, consider a computation task \(x\) that has data stored on level $j$, according to \cref{eq:mr_ji} and \cref{eq:P_i}, when \(P_x^j=\min(P_{\text{peak}}^{j}, B_{\text{peak}}^{j} \times I_x^j) \geq B_{\text{peak}}^{j, i} \times I_x^j\), we have \(P_x^i \leq B_{\text{peak}}^{j, i} \times I_x^j \leq P_x^j\). Therefore, the first turning point $P_1$ is at:
\begin{equation}
\bar{I}_x^{j} = \frac{\min(P_{\text{peak}}^{j}, B_{\text{peak}}^{j} \times I_x^j)}{B_{\text{peak}}^{j, i}} \label{eq:inter1}
\end{equation}
This gives the critical operational intensity \(\bar{I}_x^{j}\), indicating the threshold below which it is not beneficial to transfer data from level $j$ to $i$ for computation for \(x\).

Now if we continue increasing \(I_x^{j}\) such that \(P_x^j < B_{\text{peak}}^{j, i} \times I_x^j \leq \min(P_{\text{peak}}^{i}, B_{\text{peak}}^{i}\times I_x^i)\), then we obtain another turning point $P_2$:
\begin{equation}
\bar{I}_x^{j} = \frac{\min(P_{\text{peak}}^{i}, B_{\text{peak}}^{i}\times I_x^i)}{B_{\text{peak}}^{j, i}}\label{eq:inter2}
\end{equation}
which denotes the critical operational intensity \(\bar{I}_x^{j}\) below which computation \(x\) is bounded by the memory bandwidth from memory at level $j$ to memory at level $i$.

\paragraph{Balance Point}
Further, if \(B_{\text{peak}}^{i}\times I_x^i < B_{\text{peak}}^{j, i} \times I_x^j< P_{\text{peak}}^i\), indicating that the computation \(x\) on level $i$ is memory-bound (refer to \cref{eq:tp1}). In this situation, further increasing \(I_x^{j}\) cannot improve the system's performance. Instead, we need to increase \(I_x^{i}\), and a balance point will be reached if:
\begin{equation}
 B_{\text{peak}}^{i}\times I_x^i = B_{\text{peak}}^{j, i} \times I_x^j \label{eq:balance}
\end{equation}
Our performance model and policy optimizer (see \cref{sec:performance_model}) are designed to find the maximum balance point under the device memory constraints.

\vspace{-1em}
\begin{figure}[ht]
    \centering
\includegraphics[width=0.85\linewidth]{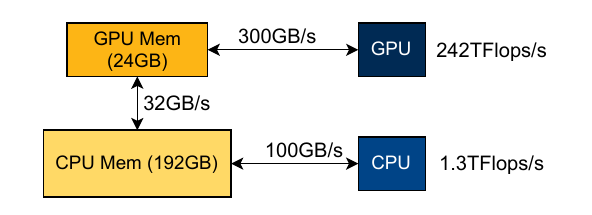}
    \vspace{-1em}
    \caption{Hardware Configurations for the L4 Instance.}
    \vspace{-1em}
    \label{fig:hardware}
\end{figure}

\subsection{Case Study} \label{sec:case_study}
To visualize the turning points and balance points discussed in the preceding sections, we conduct a case study with real HRM plots for computations\footnote{We only discuss the attention and feed-forward blocks since they account for the majority of computation time and represent quite different computation characteristics.} in a single layer of the Mixtral 8x7B model on a Google Cloud Platform L4 instance. The hardware setting is as detailed in \cref{fig:hardware}. Specifically, we let levels \(i\) and \(j\) represent GPU and CPU, respectively. Then, we define the following:

\begin{definition}[Batch Size $N$]
Batch size is the total number of tokens processed by one pass of the whole model. 
\end{definition}

\begin{definition}[Micro-Batch Size $\mu$]
Since GPU memory is limited, a batch of size $N$ often needs to be split into several micro-batches of size $\mu$ to be processed by a single kernel execution on GPU.
\end{definition}
\vspace{-1em}
\begin{figure}[ht]
    \centering
    \includegraphics[width=0.82\linewidth]{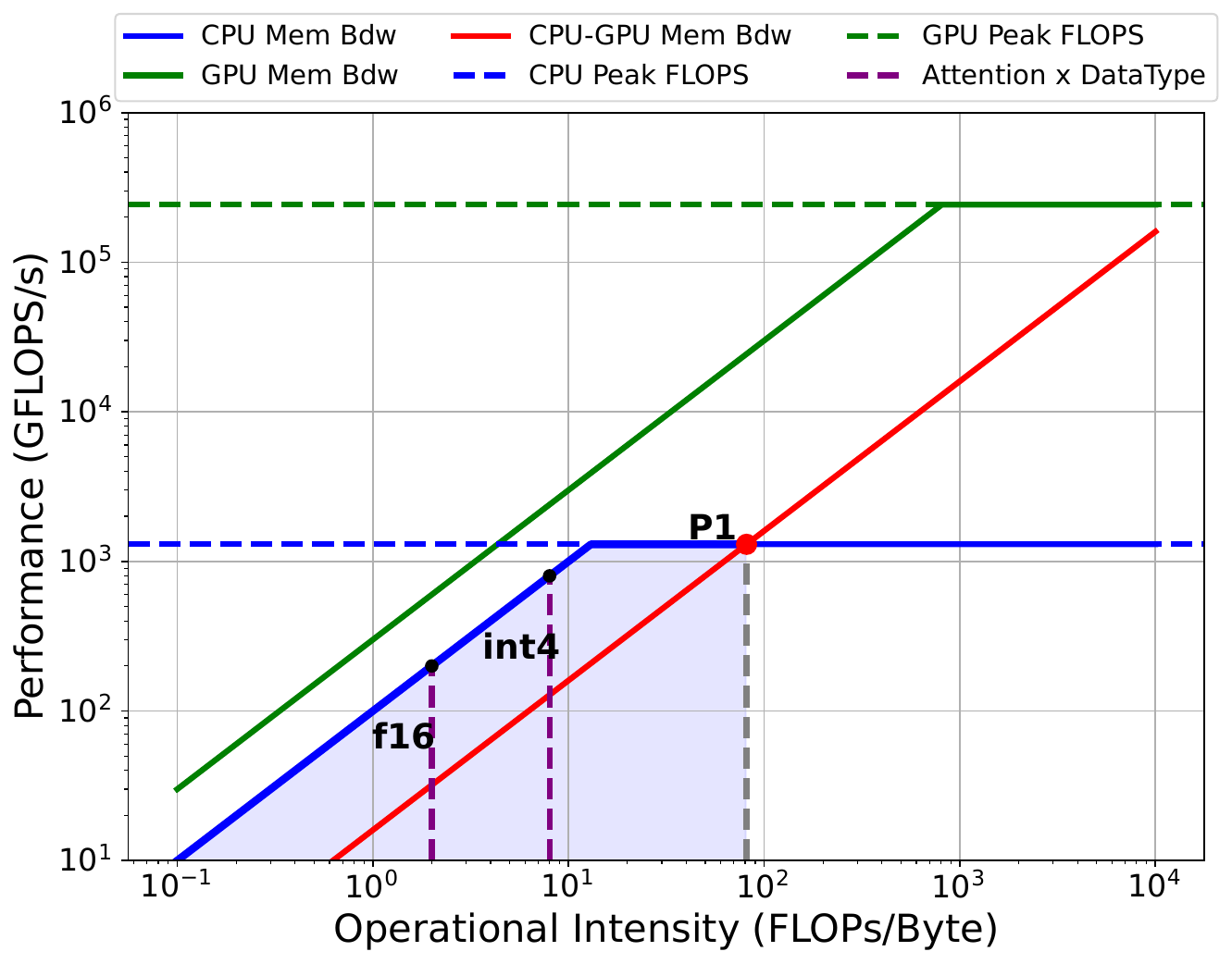}
    \vspace{-0.5em}
    \caption{Hierarchical Roofline Model for Mixtral 8x7B's Grouped Query Attention Block in Decode Stage on L4 Instance. (Context Length = 512)}
    \vspace{-1em}
    \label{fig:roofline-attn}
\end{figure}

\paragraph{Attention Block} \cref{fig:roofline-attn} demonstrates the HRM plot for Mixtral 8x7B's attention computation\footnote{Not including QKVO projection.} assuming all the KV cache are stored on CPU\footnote{For analysis purposes, we use the calculated theoretical operational intensity instead of numbers from real profiling}. 
On the plot, we have horizontal lines as the compute roofs defined by CPU and GPU peak performance. There are also the memory roofs defined by CPU memory bandwidth, GPU memory bandwidth, and CPU to GPU memory bandwidth, respectively. We then draw vertical lines representing different operational intensities for the attention computation with different KV cache data types.
Theoretically, attention's operational intensity is independent of the batch size since its flops and bytes are proportional to batch size. To increase the attention computation's operational intensity, we need methods such as quantization~\cite{lin2024qserve, kv_int4}, Grouped Query Attention (GQA)~\cite{ainslie2023gqa}, or sparse attention~\cite{child2019generating}. All these methods try to reduce the memory access needed by performing the attention computation, and GQA is used by most of the existing MoE models; however, as denoted in the plot, for both \texttt{float16} and \texttt{int4}\footnote{The computation is still done in \texttt{float32}.} the operational intensity is quite low and is smaller than \(P_1\)'s corresponding operational intensity, which suggests it may be better to perform attention on CPU.

\begin{figure}[ht]
    \centering
    \includegraphics[width=0.82\linewidth]{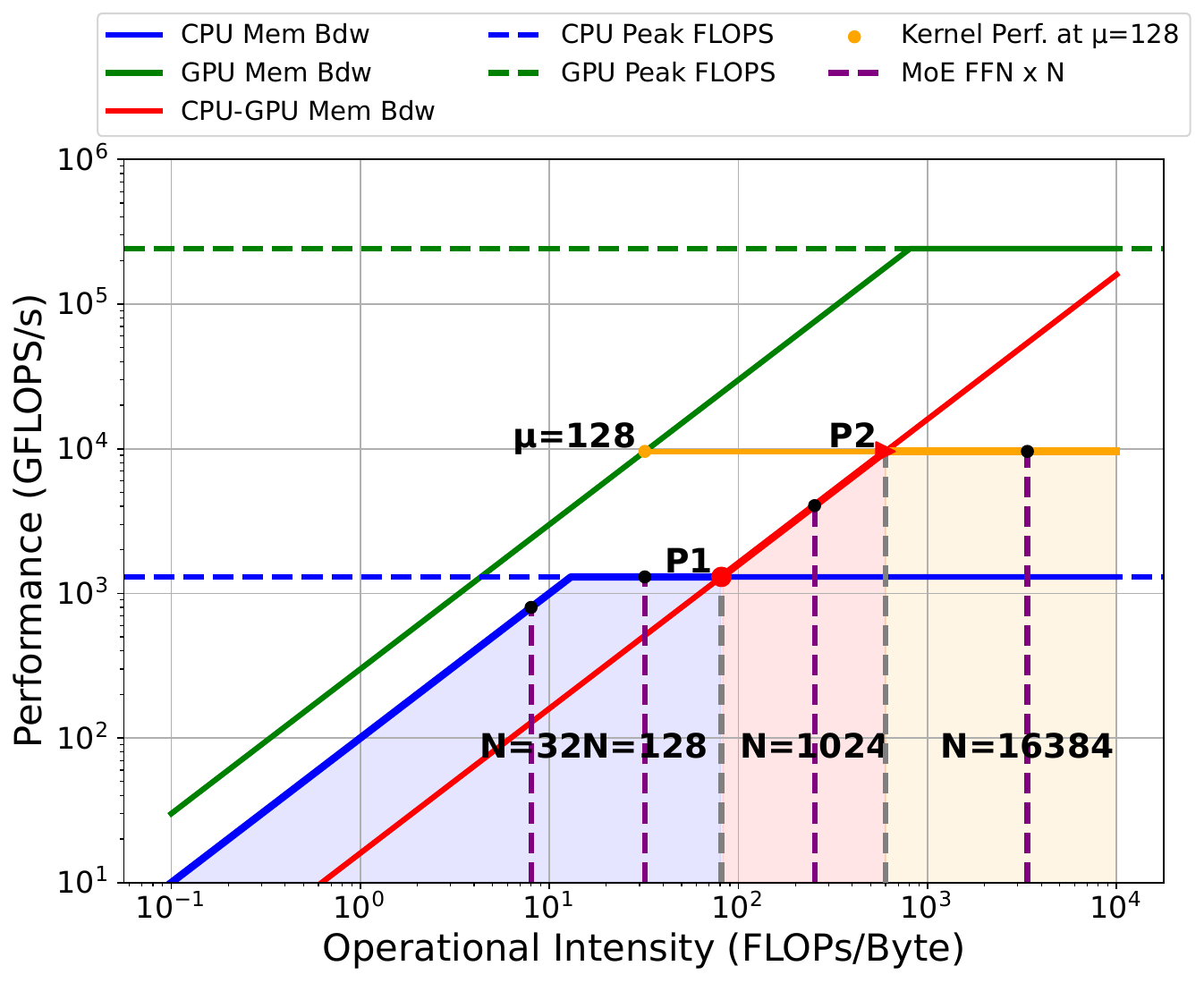}
    \vspace{-0.5em}
    \caption{Hierarchical Roofline Model for Mixtral 8x7B's MoE Feed-Forward Block in Decode Stage on L4 Instance.}
    \label{fig:roofline-mlp}
\end{figure}

\paragraph{MoE Feed-Forward Network (FFN)}
\cref{fig:roofline-mlp} is an HRM plot for Mixtral 8x7B's MoE Feed-Forward module on the L4 instance. The orange line represents the MoE FFN kernel performance achieved at a micro-batch size of 128. Vertical lines intersecting with CPU roofs and CPU-GPU memory roofs represent different batch sizes. FFN's operational intensity will increase as batch size or micro-batch size increases since, intuitively, a larger batch size means more computation per weight access. As shown in the plot, suppose the computation kernel for the MoE FFN can run at a maximum \(\mu=128\), we can identify the turning point in \cref{eq:inter2} to be \(P_2\) and the turning point in \cref{eq:inter1} to be \(P_1\). 

When \(I\) is less than \(P_1\)'s corresponding \(I\), there is no benefit in swapping the data to GPU for computation since it will be bounded by the memory roof from CPU to GPU. This is normally the case for many latency-oriented applications where users may only have one or two prompts to be processed. In such scenarios, it is more beneficial to have a static weights placement strategy (e.g., putting \(m\) out of \(n\) layers on GPU) and perform the computation where the data is located instead of swapping the weights back and forth.

Next, we show the peak performance will be finally reached at a balance point (\cref{eq:balance}).
When \(I\) is less than \(P_2\)'s corresponding \(I\), the computation is bounded by the CPU to GPU memory bandwidth, and it cannot achieve the performance at \(P_2\). Depending on whether there is enough CPU memory to hold a larger batch, we can either increase the batch size or put some of the weights on the GPU statically since both strategies can increase the operational intensity for the MoE FFN computation regarding the data on the CPU.

If the batch size can be continually increased, then when \(I\) equals \(P_2\)'s corresponding \(I\), the maximum performance that can be achieved is bounded by the operator's operational intensity on GPU, which is dependent on the \(\mu\) for the MoE FFN kernels. Then, there is no need to increase \(N\) anymore, and the maximum performance reached at a balance point equals \(P_2\). 
On the other hand, if we put more weights onto GPU, \(\mu\) will decrease since larger \(\mu\) will result in higher peak memory consumption. The maximum performance will be achieved at a balance point smaller than \(P_2\).

In conclusion, to achieve high throughput for batched MoE inference, we hope to place computations on proper computing devices and find the best combination of \(N\) and \(\mu\) so that we can fully utilize all the system's components.

\section{Method}
In general, we adopt the zigzag computation order proposed in FlexGen~\cite{sheng2023flexgen}: loading the weights from CPU\footnote{We do not consider disk offloading in this work.} and performing the computation layer by layer. For the prefill stage, we perform all the computation on GPU and offload KV cache to CPU for all the micro-batches\footnote{Since the prefill stage is normally compute-bound, and the computation can be easily overlapped with I/O, we do not perform further optimization for prefill stage.}. For the decode stage, within each layer, we propose a fine-grained GPU-CPU-I/O pipeline schedule (\cref{sec:pipeline}) to increase the utilization of GPU, CPU, and I/O in \emph{decode} stage. We also build a performance model (\cref{sec:performance_model}) based on the HRM we extended from the Roofline Model to help search for the best hyper-parameters for the pipeline schedule, including the assignment of devices to perform different computations, the batch size, the micro-batch size and the ratio of weights to be placed on GPU statically. Note that for the memory-constrained scenarios we target in this paper, CPU attention is consistently better than GPU attention, according to our performance model. We also conduct an ablation study in \cref{sec:ablation2} to show how best policy changes under different hardware configurations.

\begin{figure*}[ht]
    \centering
    \includegraphics[width=0.96\textwidth]{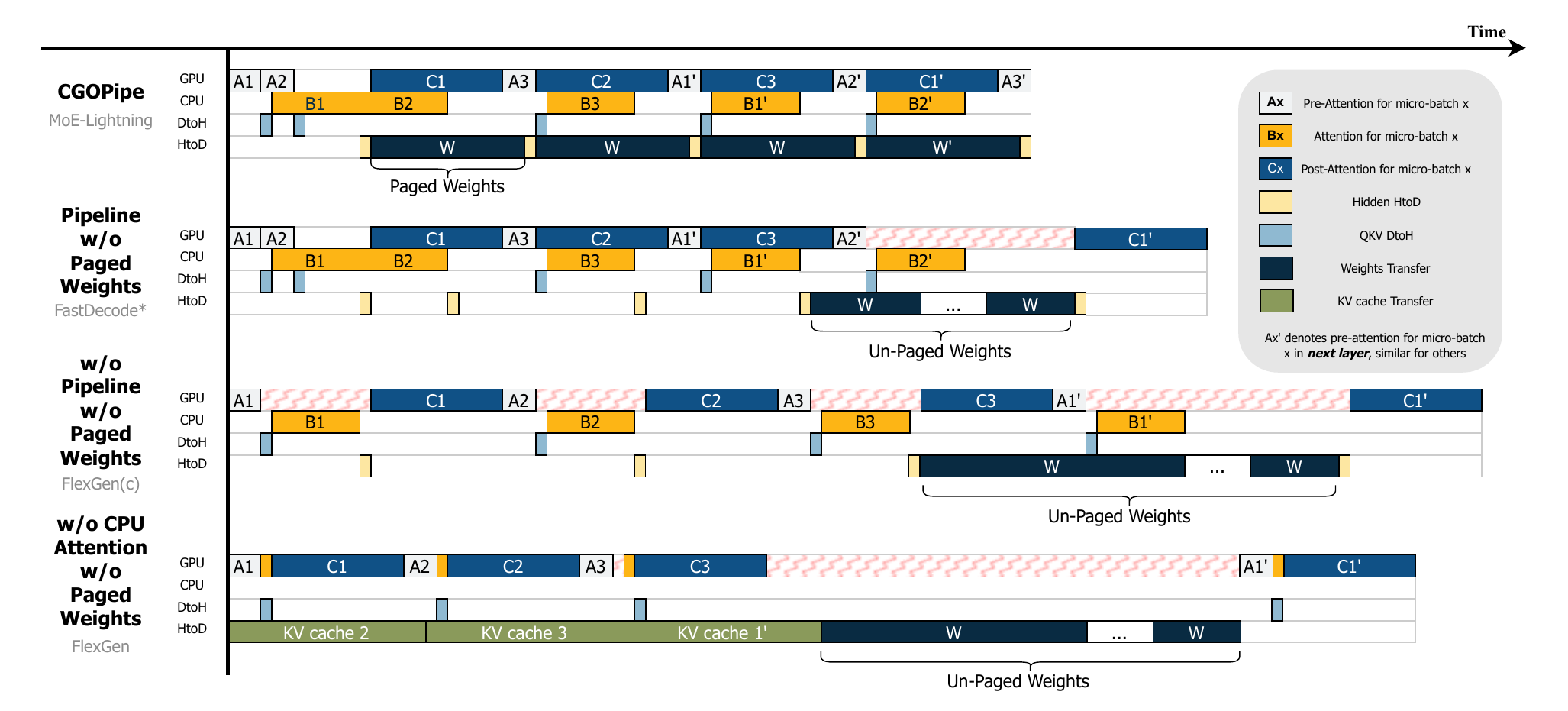}
    \vspace{-1.5em}
    \caption{Different Scheduling Strategies: Square sizes vary with workloads and policies. For example, larger $\mu$ or longer sequences lengthen the orange (attention) and the green (KV cache transfer from CPU to GPU) squares. Squares with red zigzag lines indicate the unnecessary GPU idle times. *FastDecode~\cite{he2024fastdecode} dose not consider weights offloading.}
    \label{fig:schedule}
\end{figure*}

\subsection{GPU-CPU-I/O Pipeline Schedule}\label{sec:pipeline}
\begin{algorithm}[ht]
\footnotesize
\caption{CGOPipe}
\begin{algorithmic}[1]

\For{$d = 1, 2, \ldots gen\_len$}
\State // Prologue
\For{$j = 1, 2$}
\State \Call{PreAttn}{$1, j$}
\State \Call{OffloadQKV}{$1, j$}
\State \Call{CPUAttn}{$1, j$}
\State \Call{W\_CtoPin}{$2, j$}
\EndFor

\For{$i = 1, 2, \ldots num\_layers$}
\For{$j = 1, 2, \ldots num\_ubs $}
\State \Call{LoadH}{$i, j$}
\State \Call{W\_PintoG}{$i+1, j$}
\State \Call{PostAttn}{$i, j$}
\State // Launch CPUAttn two batches ahead
\State \Call{PreAttn}{$i, j+2$}
\State \Call{OffloadQKV}{$i, j+2$}
\State \Call{CPUAttn}{$i, j+2$}
\State \Call{W\_CtoPin}{$i+1, j+2$}

\EndFor
\EndFor
\EndFor
\end{algorithmic}
\label{alg:cgopipe}
\end{algorithm}
Pipeline scheduling is a common approach to maximize compute and I/O resource utilization. Yet, the pipeline concerning GPU, CPU, and I/O is not trivial. In traditional pipeline parallelism for deep learning training~\cite{huang2019gpipe, narayanan2019pipedream, fan2021dapple}, models are divided into stages which are assigned to different devices. Therefore, only output activations are transferred between stages, resulting in a single type of data transfer in each direction at a time.
In our scenario, both weights and intermediate results need to be transferred between GPU and CPU. Intermediate results are required immediately after computation to avoid blocking subsequent operations, whereas weights for the next layer are needed only after all micro-batches for the current layer are processed. Additionally, weight transfers typically take significantly longer than intermediate results. Consequently, naive scheduling of I/O events can lead to low I/O utilization, which also hinders computation.

\noindent\textbf{\pipe.}
\cref{fig:schedule} demonstrates our proposed \pipe and the other three scheduling strategies adopted in existing systems. 
\pipe employs CPU attention as analyzed in \cref{sec:case_study}, alongside a weight paging scheme that interleaves the transfer of intermediate results for upcoming micro-batches with paged weight transfers to optimize computation and communication overlap.
The GPU sequentially processes the post-attention tasks (primarily O projection and MoE FFN) for the current micro-batch, followed by the pre-attention tasks (mainly layer norm and QKV projection) for the next micro-batch.
Concurrently, the CPU handles attention (specifically the softmax part) for the next batch, and a \emph{page} of weights for the subsequent layer are transferred to the GPU.

FlexGen~\cite{sheng2023flexgen} primarily employs the fourth schedule ($\mathcal{S}_4$), where attention is performed on GPU and the KV cache for the next micro-batch is prefetched during the current computation. This approach results in higher KV cache transfer latency than performing attention directly on the CPU (\cref{sec:case_study}) and consumes I/O bandwidth that could otherwise be used for weight transfers, reducing resource utilization compared to \pipe. 
FlexGen also supports CPU attention and adopts the third schedule ($\mathcal{S}_3$), which is the least optimized and may even perform worse than $\mathcal{S}_4$ if KV cache transfer latency is less than the sum of pre-attention, post-attention, and CPU attention latencies, as later shown by our evaluation results (\cref{sec:eval}).
FastDecode~\cite{he2024fastdecode} suggests overlapping CPU attention with GPU computation, similar to the second schedule ($\mathcal{S}_2$). However, it does not target memory-constrained settings, so weight transfer scheduling is not considered.

\noindent\textbf{Weights Paging and Data Transfer Scheduling.} To fully utilize the I/O, we propose a weights paging scheme to interleave the data transfer for different tasks, reducing bubbles in the I/O. There are mainly four kinds of data transfer: 
\begin{itemize}
\item $\mathcal{D}_1$ (QKV DtoH): the intermediate results to be transferred from GPU to CPU after QKV projection.
\item $\mathcal{D}_2$ (Hidden HtoD): the hidden states to be transferred from CPU to GPU after the CPU attention.
\item $\mathcal{D}_3$ (Weights Transfer): the weights for the next layer to be transferred from CPU to GPU.
\item $\mathcal{D}_4$ (KV cache Transfer): the KV cache for the next micro-batch to be transferred from CPU to GPU.
\end{itemize}

Due to independent data paths, data transfers in opposite directions can happen simultaneously. Data transfer will be performed sequentially in the same direction. The challenge then mainly lies in the scheduling of $\mathcal{D}_2$,  $\mathcal{D}_3$ and $\mathcal{D}_4$, which are all from CPU to GPU. For the case without CPU attention ($\mathcal{S}_4$), while $\mathcal{D}_4$ usually takes a similar or longer time compared with a layer's computation, the I/O bandwidth is almost fully utilized, leaving little room for more efficient scheduling for data transfer.
As we can see from the diagram of $\mathcal{S}_2$ and $\mathcal{S}_3$, conducting the weights transfer as a whole will block the next layer's first $\mathcal{D}_2$ for a long time, resulting in poor overall system efficiency. Instead, we can chunk the weights to be transferred into \(n\) pages where \(n\) equals the number of micro-batches in the pipeline, and the performance model and optimizer (\cref{sec:performance_model}) select the proper micro-batch size, batch size and the proportion of weights to be transferred from CPU to GPU.

Algorithm~\ref{alg:cgopipe} provides the order in which the main CPU task launcher thread launches the tasks to enable \pipe. All the tasks are executed asynchronously, and necessary synchronization primitives are added to each task to enforce the correct data dependency.
\subsection{Search Space and Performance Model}\label{sec:performance_model}

\begin{table}[ht]
\footnotesize
\centering
\caption{Notations for the Performance Model Configuration}
\vspace{-1em}
\begin{tabular}{c|c}
\hline
\textbf{Notation} & \textbf{Description} \\
\hline
\multicolumn{2}{c}{Hardware Configurations, \( \mathcal{H} \)} \\
\hline
\( m_g, m_c \) & GPU, CPU memory \\
\( b_g, b_c, b_{cg} \) & GPU, CPU, CPU-GPU bandwidth \\
\( p_g, p_c \) & GPU, CPU FLOPS \\
\hline
\multicolumn{2}{c}{Model Configurations, \( \mathcal{M} \)} \\
\hline
\( l \) & Number of layers \\
\( h_1, h_2 \) & Model, Intermediate hidden dimensions \\
\( n_q, n_{kv} \) & Query, Key/Value heads in attention \\
\( n_e, k \) & Number of experts, Top-k routing \\
\( dt \) & Data type (e.g., float32) \\
\hline
\multicolumn{2}{c}{Workload Configurations, \( \mathcal{W} \)} \\
\hline
\( s \) & Average Prompt Length \\
\( n \) & Generation Length \\
\hline
\multicolumn{2}{c}{Policy, \( \mathcal{P} \)} \\
\hline
\( N, \mu \) & Batch, Micro Batch Size \\
\(F_g, A_g\) & GPU Attention/MoE FFN Indicator \\
\(r_w, r_c\) & Ratio of Weights/KV Cache Stored on GPU \\
\hline
\end{tabular}
\label{tab:notations}
\end{table}

Given a hardware configuration $\mathcal{H}$, a model configuration $ \mathcal{M}$, and a workload configuration $ \mathcal{W}$, we search for the optimal policy $\mathcal{P}$ that minimizes per-layer latency $T(\mathcal{M}, \mathcal{H}, \mathcal{W}, \mathcal{P})$ for the pipeline schedule in \cref{sec:pipeline}, without violating the CPU and GPU memory constraints, in order to reach the optimal balance point (\cref{eq:balance}). Compared with FlexGen, we exclude disk-related variables from the search space and add two binaries to indicate whether to perform attention or MoE FFN on GPU. 

The search space (\cref{tab:notations}) covers 2 integer values: the micro-batch size ($\mu$) and batch size (\(N\)), 2 binary indicators $A_g$ to indicate whether to perform the attention on GPU and $F_g$ to indicate whether to perform the MoE FFN on GPU. When $F_g=1$, we also need to decide the percent of weights $r_w$ that can be statically stored on GPU and the percent of weights $1-r_w$ that need to be transferred to GPU. Similarly, for $A_g=1$, we need to decide $r_c$. The generated policy will be a 6-tuple $(N, \mu, A_g, F_g, r_w, r_c)$.
For our major setting, we always get $A_g=0$ and $F_g=1$. However, we discuss in \cref{sec:ablation2} different policies for various hardware settings. Notably, \pipe is primarily designed for $A_g=0$ and when $A_g=1$, \sys adopt $\mathcal{S}_4$.

We then build the performance model based on \cref{eq:P_i_j} and \cref{eq:P_i} in HRM to estimate per-layer decode latency $T$ by:
\begin{align}
    T(\mathcal{M}, \mathcal{H}, \mathcal{W}, \mathcal{P}) = \max(comm^{cpu\_to\_gpu}, T_{cpu}, T_{gpu}) \label{eq:T}
\end{align}
 
where \(comm^{cpu\_to\_gpu}\) can be computed as the number of bytes needed to be transferred from CPU to GPU for a layer's computation divided by the CPU to GPU memory bandwidth $b_{cg}$. Here, for simplicity, we only consider the attention computation and the MoE FFN computation in a transformer block, and therefore we have: 
\begin{align}
    T_{gpu} &= T_{attn}^{g} + T_{ffn}^{g} \ \text{,} \ T_{cpu} = T_{attn}^{c} + T_{ffn}^{c}
\end{align}
To estimate the time to perform a computation $x$ on GPU or CPU, we can use $T_{x} = \max(comm_{x}, comp_{x})$ according to \cref{eq:P_i} in HRM, resulting in:
\begin{align}
    T_{ffn}^{g} = \max(comm_{ffn}^{g}, comp_{ffn}^{g})
\end{align}
and similarly for $T_{attn}^{g}$, $T_{attn}^{c}$ and $T_{ffn}^{c}$.

For a given computation $x$, we can calculate their theoretical FLOPS and data transfer based on $\mathcal{M}$ and then we have $comm_{x}^g=bytes_x/b_g$ and $comp_{x}^g=flops_x/p_g$ (same for CPU). 
While there are discrepancies between the theoretical performance estimation and the kernel's real performance, such modeling can provide a reasonable estimation of the relative effectiveness of any two policies.
In this paper, all the evaluation results of \sys follow policies generated by a performance model with theoretically calculated computation flops and bytes with profiled peak performance and memory bandwidth for the hardware.

\subsection{Tensor Parallelism}
In existing works~\cite{sheng2023flexgen}, pipeline parallelism is used for scaling beyond a single GPU, which requires the number of devices to scale with the model depth instead of the layer size. However, according to our analysis for MoE models in \cref{sec:case_study}, Total GPU memory capacity can decide the upper bound throughput the system can achieve. Therefore, \sys implements tensor parallelism~\cite{narayanan2021efficient} within a single node to get a higher throughput upper bound. In this case, we have \(tp\_size\) times more GPU memory capacity and GPU memory bandwidth, we can then search for the policy similarly as for single GPU.
\section{Evaluation}\label{sec:eval}

\subsection{Setup}
\begin{table}[ht]
\centering
\footnotesize
\caption{Model and Hardware Configurations.}
\vspace{-1em}
\begin{tabular}{cccl}
\toprule
Setting & Model & GPU & CPU (Intel Xeon)\\
\midrule
S1 & Mixtral 8x7B  & 1xT4 (16G) &2.30GHz, 24-core, 192GB \\
S2 & Mixtral 8x7B  & 1xL4 (24G) &2.20GHz, 24-core, 192GB\\

\midrule
S6 & Mixtral 8x22B & 2xT4 (32G) & 2.30GHz, 32-core, 416GB \\
S7 & Mixtral 8x22B & 4xT4 (64G) & 2.30GHz, 32-core, 416GB \\
\midrule
S8 & DBRX & 2xT4 (32G) & 2.30GHz, 32-core, 416GB \\
S9 & DBRX & 4xT4 (64G) & 2.30GHz, 32-core, 416GB \\
\bottomrule
\end{tabular}
\label{tab:model_setting}
\end{table}

\begin{table}[ht]
\centering
\footnotesize
\caption{Workload Configurations.}
\vspace{-1em}
\begin{tabular}{cccl}
\toprule
Dataset & \( s_{\text{avg}} \) & \( s_{\text{max}} \) & \( l \) \\
\midrule
MTBench~\cite{zheng2024judging} & 77 & 418 & {32, 64, 128, 256} \\
Synthetic Reasoning~\cite{liang2022holistic} & 242 & 256 & 50 \\
Summarization~\cite{liang2022holistic} & 1693 & 1984 & 64 \\ 
\bottomrule
\end{tabular}
\label{tab:dataset_info}
\end{table}

\begin{figure*}[ht]
    \centering
    \includegraphics[width=0.9\textwidth]{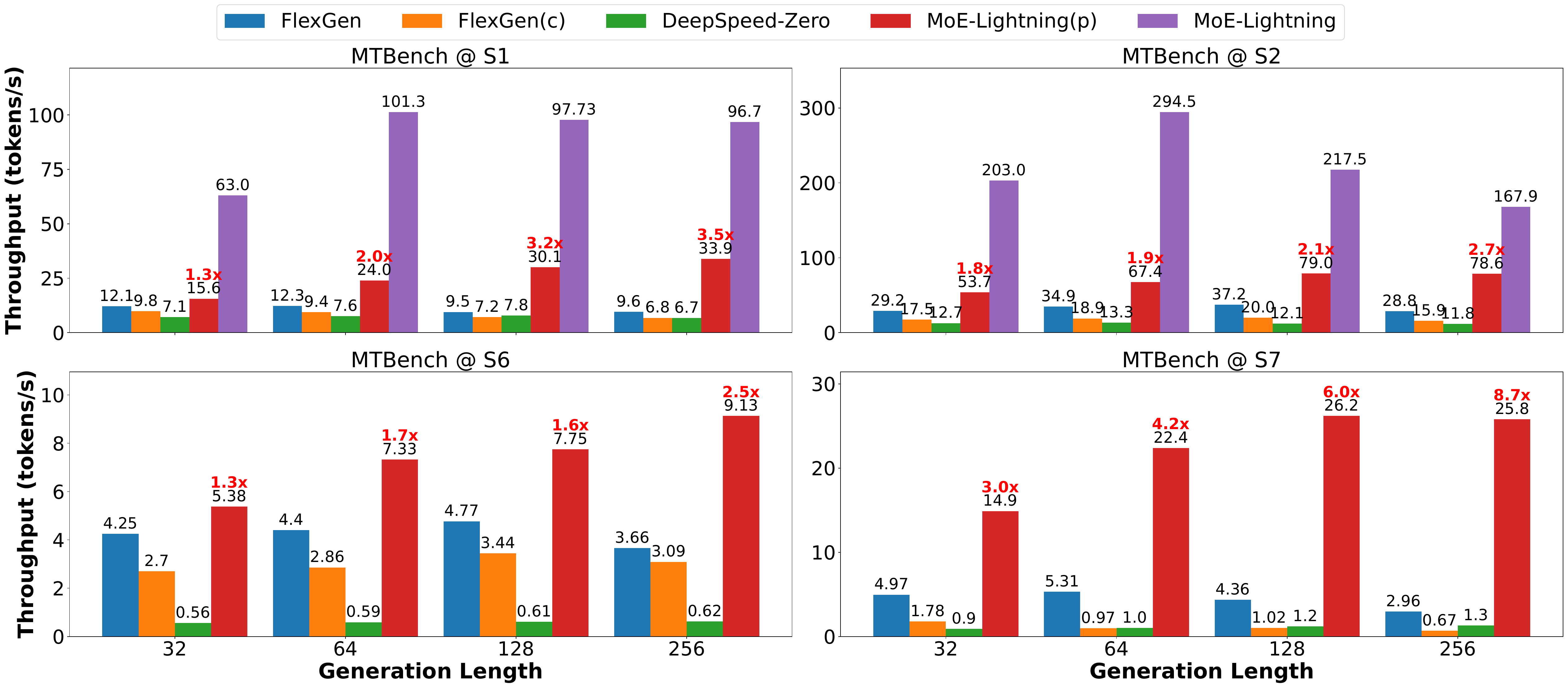}
    \caption{End-to-end Results for MTBench on Different Model-Hardware Configurations. Normally, \sys's performance will be much higher than \sys(p) since padding will lead to higher memory consumption and attention computation overhead\protect\footnotemark. Here \sys achieves up to 10.3$\times$ higher throughput than FlexGen under S1 and S2.}
    \label{fig:mtbench_all}
\end{figure*}

\begin{table*}[ht]
\centering
\footnotesize
\caption{Performance for HELM tasks under S1 \& S2 }
\vspace{-1em}
\begin{tabular}{ccccccccccccccl}
\toprule
& \multicolumn{6}{c}{\textbf{Synthetic Reasoning}} & \multicolumn{6}{c}{\textbf{Summarization}} \\
\cmidrule(lr){2-7}\cmidrule(l){8-13}
Settings & \multicolumn{3}{c}{S1} & \multicolumn{3}{c}{S2} & \multicolumn{3}{c}{S1} & \multicolumn{3}{c}{S2} \\ \cmidrule(lr){2-4}\cmidrule(lr){5-7}\cmidrule(lr){8-10}\cmidrule(l){11-13}
& Throughput & $\mu$ & $N/\mu$ & Throughput & $\mu$ & $N/\mu$ & Throughput & $\mu$ & $N/\mu$ & Throughput & $\mu$ & $N/\mu$ \\ 
\midrule
FlexGen(c) & 16.903 & 32 & 61 & 20.015 & 64 & 33 & 2.614 & 3 & 92 & 4.307 & 8 & 36 \\ 
FlexGen & 22.691 & 32 & 61 & 50.138 & 64 & 33 & 3.868 & 3 & 92 & 7.14 & 8 & 36\\
DeepSpeed & 11.832 & 102 & 1 & 18.589 & 156 & 1 & 0.965 & 8 & 1 & 1.447 & 12 & 1 \\
\midrule
MoE-Lightning(p) & \textbf{26.349} & 36 & 26 & \textbf{105.29} & 100 & 15 & \textbf{4.52} & 4 & 19 & \textbf{12.393} & 8 & 36\\
\bottomrule
\end{tabular}
\label{tab:helm}
\end{table*}

\noindent\textbf{Implementation.} We build \sys on top of PyTorch~\cite{paszke2019pytorch}, vLLM~\cite{kwon2023efficient} and SGLang~\cite{zheng2024sglang}, written in Python and C++. We implement customized CPU Grouped Query Attention (GQA) kernels based on Intel's MKL library~\cite{mkl}. 

\noindent\textbf{Models.} We evaluate three popular MoE models: Mixtral 8x7B~\cite{mixtral}, Mixtral 8x22B~\cite{mixtral22b}, and DBRX (132B, 16 Experts)~\cite{dbrx}. Although not evaluated, \sys also supports other models compatible with vLLM~\cite{kwon2023efficient}'s model classes. 

\noindent\textbf{Hardware.} We conduct tests on various hardware settings, including a
 single NVIDIA T4 GPU (16GB), a single NVIDIA L4 GPU (24GB) and multiple T4 GPUs. 
 We evaluate 6 different model and hardware settings as shown in \cref{tab:model_setting}. 

\noindent\textbf{Workloads.} We use popular LLM benchmarks with different prompt length distributions to evaluate our system, as shown in \cref{tab:dataset_info}. MTBench~\cite{zheng2024judging} includes 80 high-quality multi-turn questions across various categories like writing and reasoning. We replicate it into thousands of questions for our batch inference use case. We test various output token lengths for MTBench, from 32, 64, 128, to 256 tokens. We also pick two tasks (i.e., synthetic reasoning and summarization), from the HELM benchmarks~\cite{liang2022holistic} to test our system with longer prompt lengths.

\noindent\textbf{Baselines.}
We evaluate \sys and \sys's variant, comparing them against two baseline systems that support running LLMs without enough GPU memory: FlexGen~\cite{sheng2023flexgen} and DeepSpeed Zero-Inference~\cite{aminabadi2022deepspeed}.

\begin{itemize}\setlength{\itemsep}{0pt}
    \item FlexGen~\cite{sheng2023flexgen} is the state-of-the-art offloading system that targets high-throughput batch inference for OPT~\cite{opt} models. It does not support variable prompt length in a batch and needs to pad all the requests to the maximum prompt length in the batch.
    \item FlexGen(c) is FlexGen enabling CPU attention.
    \item DeepSpeed Zero-Inference~\cite{aminabadi2022deepspeed} is an offloading system that pins model weights to CPU memory and streams them layer-by-layer to GPU for computation. We use version 0.14.3 in the evaluation.
    \item \sys represents our system with all the optimizations enabled.
    \item \sys(p) represents our system running with requests padded to the maximum prompt length in the batch to compare with FlexGen. 
\end{itemize}

\textbf{Metrics.} We measure the \emph{generation throughput} for each workload, which is calculated as the number of tokens generated divided by total generation time (i.e., prefill time + decode time). 

\subsection{End-to-end Results on Real Workloads}

We evaluate the maximum generation throughput for all baseline systems on three workloads under S1, S2, S6, and S7 settings. As shown in \cref{fig:mtbench_all} and \cref{tab:helm}, \sys(p) outperforms all baselines in all settings, and \sys achieves up to $10.3\times$ better throughput compared with the best of the baselines for MTBench and HELM benchmark.
\footnotetext{We only show \sys's results for S1 and S2 and omit them for S6 and S7 to focus on the comparison of the main optimizations we proposed in this paper.}
In the following sections, we analyze how \sys(p) outperforms our baselines by integrating the key methods from \cref{sec:performance_model}.

\noindent\textbf{Generation Length.}
While longer lengths allow for better amortization of the prefill time which increases throughput, they also lead to higher CPU memory usage and additional attention computation or KV cache transfer overheads.
This increased memory demand can limit the maximum batch size, reducing throughput. Moreover, the increase in computation or KV cache transfers can make attention the main bottleneck. Typically, throughput first increases with longer generation length and then decreases.

We observe this pattern for FlexGen and FlexGen(c) in all settings.
However, \sys(p) avoids a decrease in throughput under S1 and S6, which feature similar ratios of GPU to CPU memory.
We attribute this performance improvement to \pipe, which significantly improves the resource utilization and renders the system GPU memory capacity bound in these settings.

On a single GPU (S1 and S2), \sys(p) achieves up to 3.5$\times$, 5$\times$, and 6.7$\times$ improvement over FlexGen, FlexGen(c), and DeepSpeed, respectively. 

\noindent\textbf{Prompt Length.} 
In the HELM tasks, we examine the impact of varying prompt lengths on generation throughput. Increasing the prompt length not only raises CPU memory consumption and attention overhead, but also leads to greater GPU peak memory usage during the prefill stage. Consequently, systems handling the summarization task with a 2k prompt length are bottlenecked by either GPU memory capacity or attention processes (see the ablation study in \cref{sec:ablation2} for a detailed discussion on bottlenecks). Under S1, \sys(p) achieves a 1.16$\times$ and 1.73$\times$ higher throughput than FlexGen and FlexGen(c), respectively, despite using a batch size that is 3.63$\times$ smaller, enabled by \pipe. DeepSpeed, utilizing a larger micro-batch size but the smallest batch size, is primarily constrained by the overhead of weight transfers. Under S2, with increased GPU memory, \sys(p) adjusts to use a larger $\mu$ and $N$, reaching a new balance point (\cref{eq:balance}), while FlexGen and FlexGen(c) are unable to increase $N$ from their S1 settings due to CPU memory limitations. As a result, \sys(p) now achieves an even higher throughput improvement: 1.74$\times$ and 2.88$\times$ higher than FlexGen and FlexGen(c), respectively. This superior performance is attributed to \sys(p)'s efficient resource utilization.

The synthetic reasoning task enables all systems to have a larger micro-batch size due to the shorter prompt length.
Under S1, \sys(p) achieves a 1.16$\times$, 1.56$\times$, 2.22$\times$ higher throughput than FlexGen, FlexGen(c) and DeepSpeed respectively. Under S2, \sys(p) finds a better balance point and uses less batch size than FlexGen, achieving 2.1$\times$ and 5.26$\times$ higher throughput compared to FlexGen and FlexGen(c), demonstrating the efficiency of \pipe and \cm.

\subsection{Tensor Parallelism}
This section evaluates \sys's ability to run on multiple GPUs with tensor parallelism. As shown in S1 and S2, due to our efficient resource utilization, \sys's throughput is predominantly bounded by GPU memory capacity.
This shows that increasing GPU memory can raise the system's throughput upper bound. 
\begin{figure}[ht]
    \centering
    \includegraphics[width=0.42\textwidth]{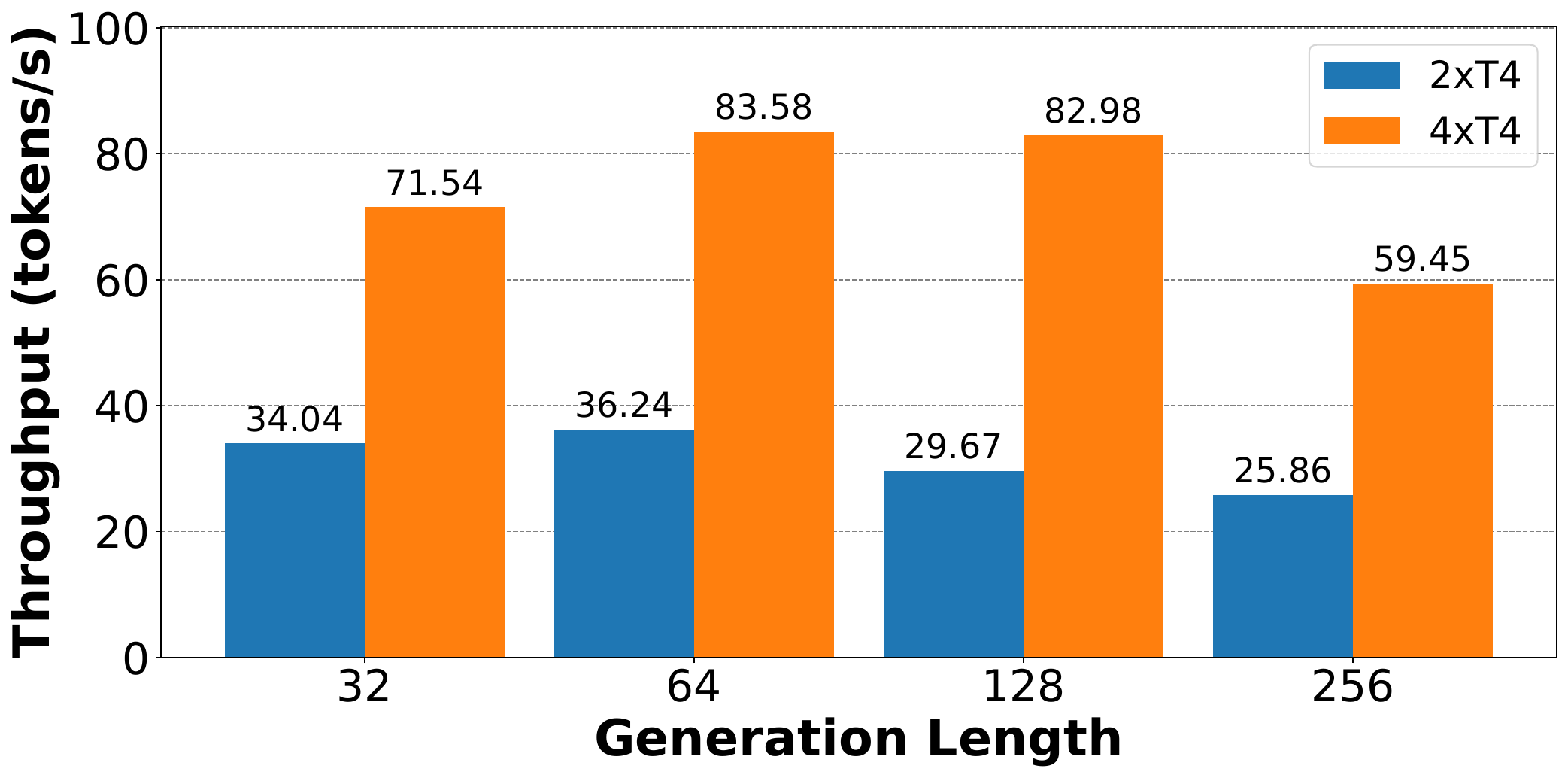}
    \vspace{-0.8em}
    \caption{\sys with Tensor-Parallelism for MTBench @ S8 \& S9.}
    \label{fig:mtbench_s8&9}
\end{figure}

S6 and S7 in \cref{fig:mtbench_all} show the end-to-end throughput results on Mixtral 8x22B of \sys(p), FlexGen, and DeepSpeed on MTBench for multiple T4 GPUs. 
Notably, \sys(p) achieves 2.77-3.38$\times$ higher throughput with 4xT4 GPUs than with 2xT4 GPUs, demonstrating super-linear scaling performance. DeepSpeed demonstrates a linear-scaling performance but uses a small batch size of 32, resulting in low throughput. FlexGen fails to scale under settings S6 and S7, largely due to the pipeline parallelism approach it employs. In this method, when using 4 GPUs, during the saturated phase, four layers are simultaneously active across four GPUs, increasing CPU peak memory consumption. As a result, FlexGen is bottlenecked by the CPU to GPU memory bandwidth and fails to take advantage of the added GPUs. Note that pipeline parallelism is more effective across multiple GPU nodes. In such configurations, doubling the number of GPUs also doubles the CPU to GPU bandwidth, the CPU memory capacity, and the CPU memory bandwidth\footnote{In this paper, we focus on the cases within one node.}.

\cref{fig:mtbench_s8&9} demonstrates \sys's generation throughput results on DBRX to showcase the performance when all optimizations are enabled (\pipe, \cm and variable length prompts). For the DBRX model and without request padding (i.e., shorter prompt length), the system becomes less GPU memory capacity bound. We can see 2.1-2.8$\times$ improvement when scaling from 2 GPUs to 4 GPUs.
\section{Ablation Study}
\subsection{Optimizer Policy}
In this section, we compare \sys(p), FlexGen with its policy and FlexGen with our policy. For this experiment, we do not turn on the CPU attention for FlexGen as it is consistently worse than FlexGen w/o CPU attention. We use the workload from MTBench on the S1 setting with a generation length of 128. The results are displayed in \cref{tab:ablation_policy}. By deploying our policy, we can see a $1.77\times$ improvement in FlexGen. We also increase the batch size to better amortize the weights transfer overhead and it gives a $2.17\times$ speedup. However, it still cannot match \sys's throughput under the same policy, as KV cache swapping becomes the bottleneck for FlexGen in this case.
\begin{table}[ht]
\centering
\footnotesize
\caption{Generation throughput for \sys and different variants of FlexGen. (MTBench@S1, Generation length=128)}
\vspace{-1em}
\begin{tabular}{cccl}
\toprule
 & $\mu$ & $N$ & Throughput (token/s)\\
\midrule
FlexGen w/ their policy & 8 & 1112 & 9.5 \\
FlexGen w/ our policy  & 36  & 504 & 16.816 ($1.77\times$) \\
FlexGen w/ our policy + larger $N$ & 36  & 1116 & 20.654 ($2.17\times$) \\
\midrule
\sys(p) & 36 & 504 & 30.12 ($3.17\times$) \\
\bottomrule
\end{tabular}
\label{tab:ablation_policy}
\end{table}

\subsection{CPU Attention vs. Experts FFN vs. KV Transfer}\label{sec:ablation_bound}
In this section, we study when CPU attention will become the bottleneck in the decode stage. For different batch sizes (from 32 to 256), we test the latency of the MoE FFN kernel on L4 GPU and compare it with the latency of the CPU attention kernel on a 24-core Intel(R) Xeon(R) CPU @ 2.20GHz with various context lengths (from 128 to 2048). Additionally, we also measure the latency for swapping the KV cache needed for the attention from CPU pinned memory to GPU to validate the efficiency of our CPU GQA kernel. 
\begin{figure}[ht]
    \centering
    \includegraphics[width=0.47\textwidth]{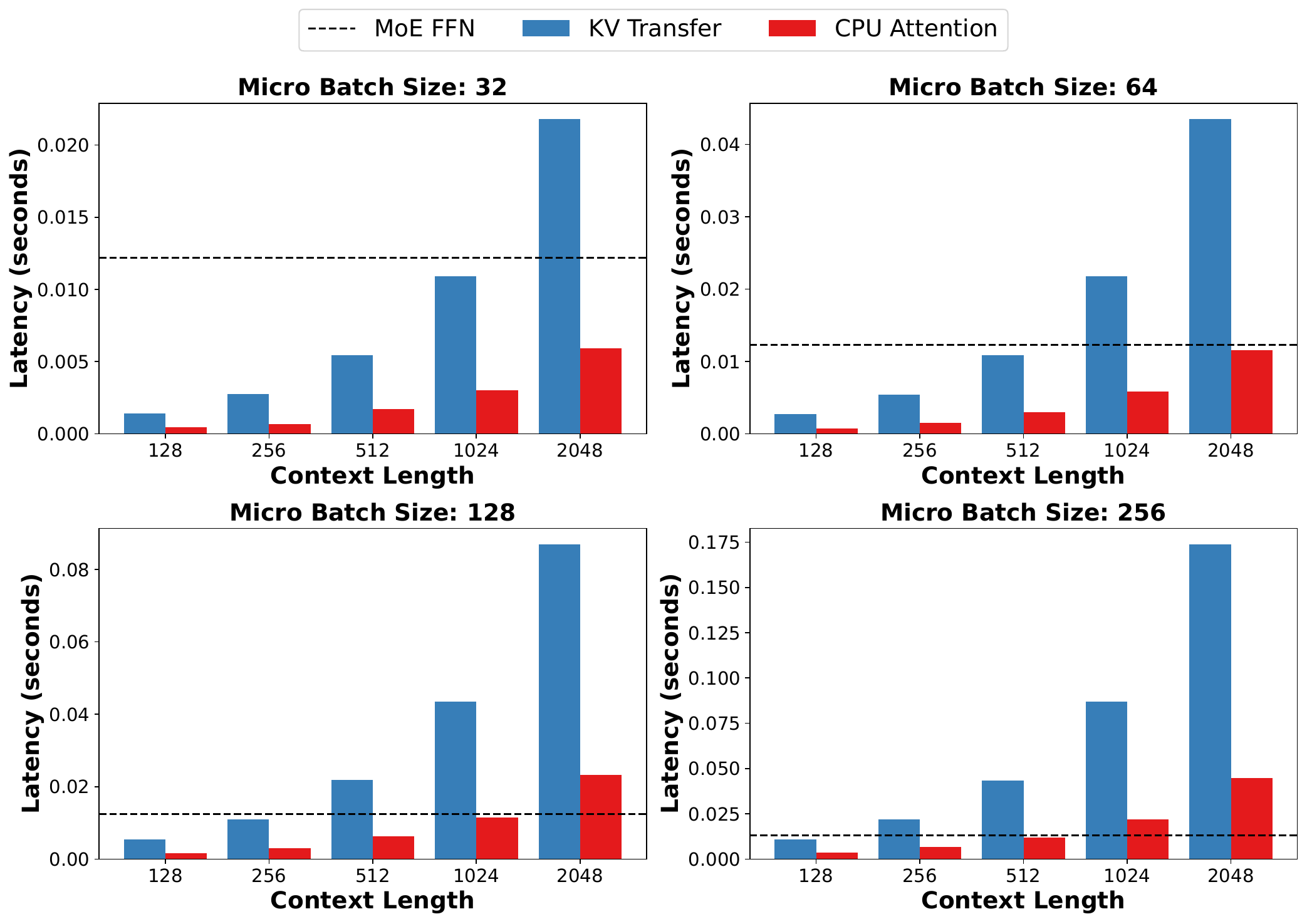}
    \caption{Latency Comparison for a single layer's KV cache transfer, CPU Attention Kernel and the MoE FFN Kernel wrt. $\mu$ and Context Length in Decode Stage.}
    \label{fig:ablation_2}
\end{figure}

As shown in \cref{fig:ablation_2}, our CPU attention kernel is \(3-4\times\) faster than KV cache transfer, which is close to the ratio of CPU memory bandwidth and the CPU to GPU memory bandwidth. The MoE FFN's latency doesn't change so much across different micro batch sizes, which is as expected since the kernel is memory-bound for the decode stage. As the micro-batch size and context length increase, the CPU attention will eventually become the bottleneck, which calls for higher CPU memory bandwidth.

\subsection{Case Study on Different Hardware Settings}\label{sec:ablation2}

In this section, we study how the best policy changes under different hardware settings. 
As we have shown in the previous ablation study, CPU attention can actually become the bottleneck for large batch size and context length, which means if we have more powerful GPUs, at some point, CPU attention may not be worth it. Moreover, if we have higher CPU to GPU memory bandwidth, the trade-offs will also change. Then the question becomes: when we have enough GPU memory (e.g., 2xA100-80G) to hold the model weights (e.g., Mixtral 8x7B), is it still beneficial to perform CPU computation or to offload weights/KV cache to the CPU? To conduct the analysis, we use 2xA100-80G for the GPU specification and vary the CPU to GPU memory bandwidth from 100 to 500 GB/s alongside different CPU capabilities. We set base CPU specifications at $m_c=200$GB/s, $b_c=100$GB, and $p_c=1.6$TFLOPS/s, scaling these values by multiplying with the CPU scaling ratio for various configurations\footnote{Note that this setup doesn't reflect real-world hardware scaling; rather, it simplifies the number of variables to offer a rough idea of how these hardware configurations might impact performance.}.
\begin{figure}[ht]
    \centering
    \includegraphics[width=0.47\textwidth]{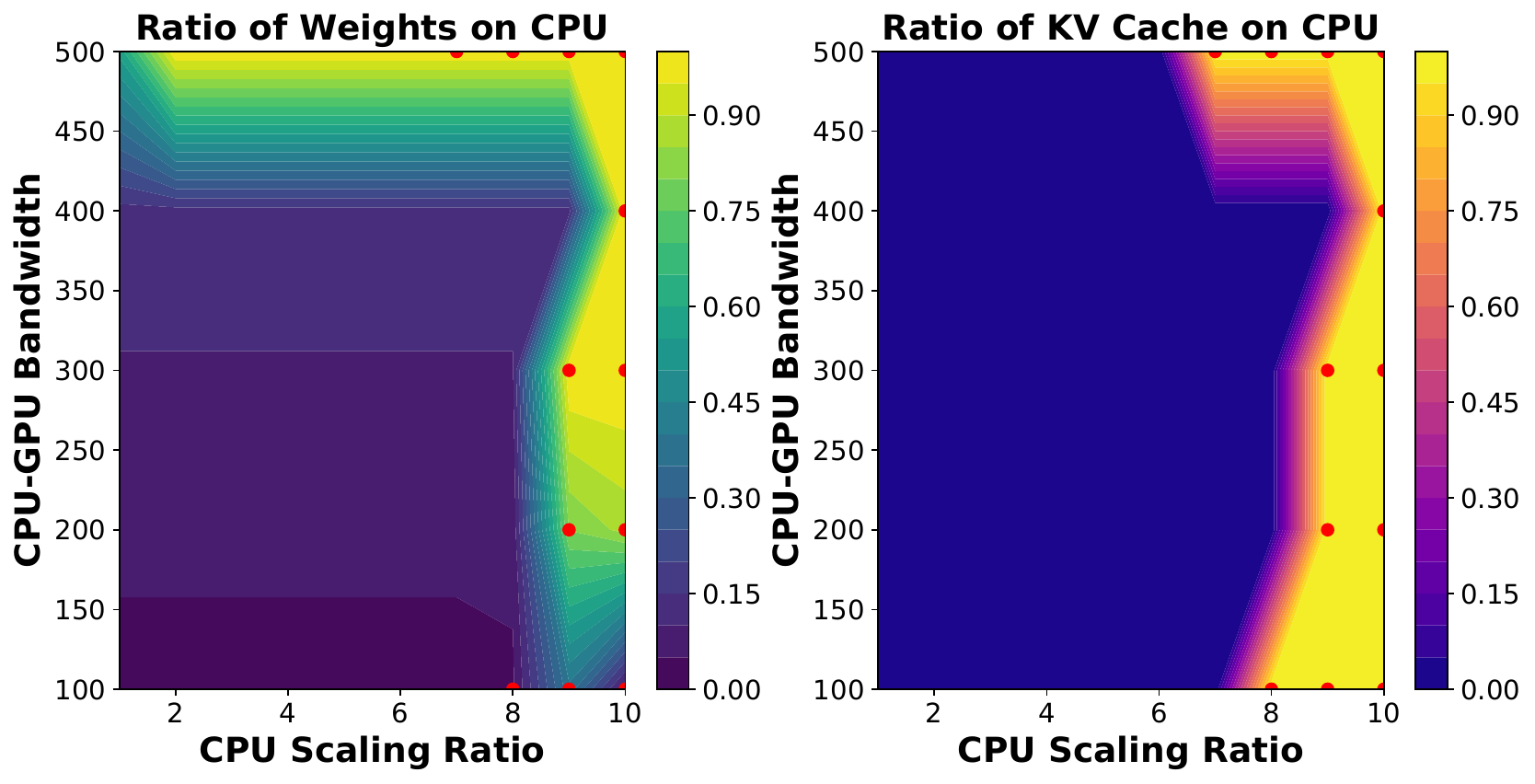}
    \caption{Policy changes with different hardware configurations (prompt length=512, generation length=32). Red points denote performing attention on the CPU.}
    \label{fig:contour}
\end{figure}

We can see that when running Mixtral 8x7B on two A100 GPUs, as CPU-to-GPU memory bandwidth increases, more weight will be offloaded to the CPU. KV cache offloading is highly related to the CPU scaling ratio in this setup: when the CPU scaling ratio is low (i.e., low CPU memory bandwidth), even with the highest CPU to GPU memory bandwidth tested here, it is not beneficial to offload KV cache.  

\section{Related Work}

\textbf{Memory-constriant LLM Inference} LLM inference requires substantial memory to store model parameters and computation outputs, making it typically memory capacity-bound. There is a line of research dedicated to memory-constraint LLM inference. This is particularly crucial for inference hardware such as desktop computers, or low-end cloud instances with limited computational power and memory capacity. To facilitate inference on such constrained systems, some work leverages sparsity or neuro activation patterns to intelligent offloading between CPU and GPU~\cite{xue2024moeinfinity, song2023powerinfer, eliseev2023fast, sheng2023flexgen}. Some approaches utilize not only DRAM but also flash memory to expand the available memory resources~\cite{alizadeh2024llm}. Additionally, since the CPU often remains underutilized during inference, it can be harnessed to perform complementary computations~\cite{song2023powerinfer, xuanlei2024hetegen, kamahori2024fiddler}.

\noindent\textbf{LLM Inference Throughput Optimization}
To enhance inference throughput, some research focuses on maximizing the sharing of computations between sequences to minimize redundant processing of identical tokens~\cite{juravsky2024hydragen, zheng2024sglang}. Another approach involves batching requests~\cite{yu2022orca} to optimize hardware utilization. Additionally, some studies develop paged memory methods for managing the key-value (KV) cache to reduce memory waste, thereby increasing the effective batch size and further improving throughput~\cite{kwon2023efficient}. FastDecode~\cite{he2024fastdecode} proposes aggregating memory and computing power of CPUs across multiple nodes to process the attention part to boost GPU throughput. Compared with FastDecode, we are targeting the memory-constrained case where the model weights also need to be transferred between CPU and GPU, making the optimization and scheduling problem far more challenging.

\noindent\textbf{LLM Inference Latency Optimization} To reduce LLM inference latency, some work addresses the inherent slowness caused by the autoregressive nature of LLM inference by developing fast decoding methods, such as speculative decoding~\cite{chen2023accelerating, stern2018blockwise, leviathan2023fast} and parallel decoding~\cite{Santilli_2023}, which generate multiple tokens simultaneously. Another approach aims to decrease inference latency by implementing efficient computational kernels~\cite{dao2022flashattention, dao2023flashattention2, flashinfer2024} designed to minimize memory access and maximize GPU utilization.
\vspace{-1em}

\section{Conclusion}
We present \sys, a high-throughput MoE inference system for GPU-constrained scenarios. \sys can achieve up to 10.3$\times$ (without request padding) and 3.5$\times$ (with request padding) higher throughput over state-of-the-art systems on a single GPU and demonstrate super-linear scaling on multiple GPUs, enabled by \pipe and \cm.
\pipe is a novel pipeline scheduling strategy to improve resource utilization, and \cm is a performance model based on a Hierarchical Roofline Model that we extend from the classical Roofline Model to find policies with higher throughput upper bound.

\appendix
\section{System Implementation Details}
In this section, we explain two system-level designs and their implementation details: 1. \cref{sec:mem} introduces how GPU and CPU memory are used and weights paging is implemented in \sys, and 2. \cref{sec:batching} presents the batching algorithm employed in \sys to support dynamic-length requests in a batch.
\subsection{Memory Management}\label{sec:mem}
Since attention is performed on CPU, the KV cache for all micro-batches will be transferred to and stored on CPU after the corresponding computation completes. To enable \pipe, we allocate a weight buffer with a size of $2\times sizeof(W_L)$, where $W_L$ denotes the size of the portion of a layer's weights stored in CPU memory. This buffer enables overlapping weight prefetching: as the current layer's weights are being used, the next layer's weights are simultaneously transferred to GPU memory.

\begin{figure}[ht]
    \centering
    \includegraphics[width=0.49\textwidth]{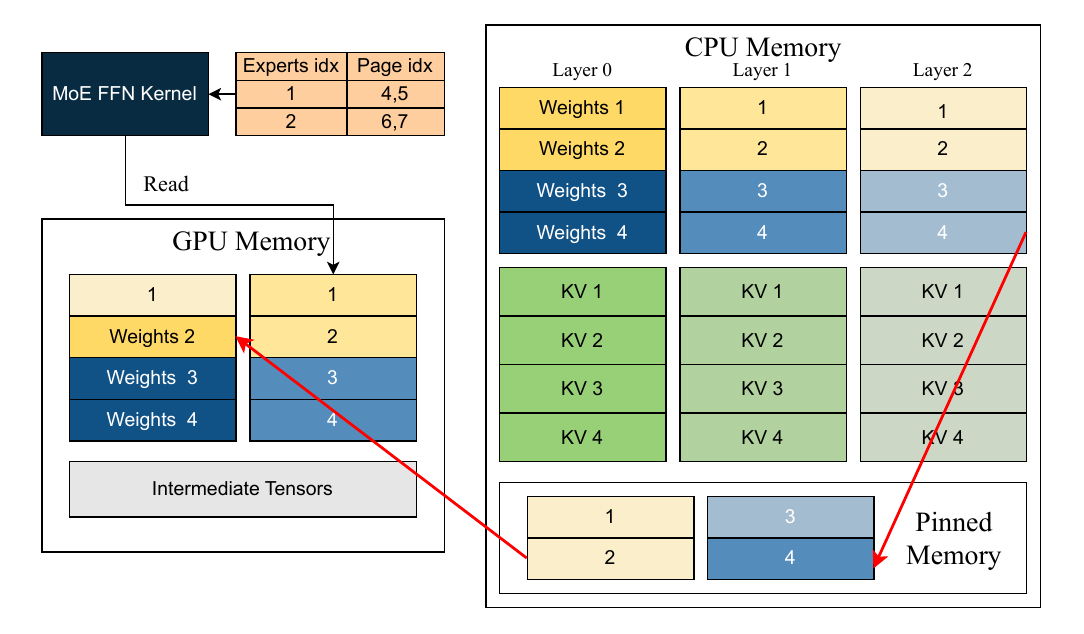}
    \caption{Simplified Demonstration of \sys's Memory Management.}
    \vspace{-1em}
    \label{fig:memory}
\end{figure}

Weights are transferred in a paged manner. For example in \cref{fig:memory}, each expert in the MoE FFN kernel requires two pages, and the kernel accesses the appropriate pages using a page table. To accelerate transfers from CPU to GPU, weights are first moved from CPU memory to pinned memory, and then from pinned memory to GPU. These transfers are overlapped to hide latency. As illustrated in \cref{fig:memory}, while transferring \texttt{Weights 2} for Layer 2 from pinned memory to GPU, \texttt{Weights 4} for the same layer can be transferred concurrently from CPU to pinned memory.

\subsection{Request Batching}\label{sec:batching}
For a given workload, the optimizer introduced in \cref{sec:performance_model} takes the average prompt length to search for an optimal policy. However, maintaining a consistent micro-batch size becomes challenging due to varying input lengths across requests. To address this, we employ the strategy outlined in \cref{alg:batching} to achieve balanced token distribution. In essence, requests are sorted by input length in descending order and assigned to micro-batches by iteratively placing the longest request into the micro-batch with the fewest tokens. This approach ensures that all micro-batches have a size close to the $ubs$ specified by the generated policy.
\begin{algorithm}[ht]
\caption{Request Batching}
\label{alg:batching}
\begin{algorithmic}[1]

\Require $req\_queue$: Queue of requests
\Require $n\_ub$: Number of micro-batches
\Require $ubs$: Maximum number of requests per micro-batches
\Require $gen\_len$: Generation length per request
\Require $cache\_size$: Maximum cache size per micro-batches
\Ensure $micro\_batches$: List of micro-batches
\Ensure $aborted\_requests$: List of aborted requests to be added to the next batch

\For{$i \gets 1$ \textbf{to} $n\_ub$}
    \State $partitions[i] \gets \emptyset$
    \State $partitions\_sums[i] \gets 0$
\EndFor

\State \Call{Sort}{$req\_queue$, $key$ = $\lambda x.\, x.input\_len$, $reverse = True$}

\ForAll{$req \in req\_queue$}
    \If{$partitions == \emptyset$}
        \State $aborted\_requests \mathrel{+}= req$
    \EndIf
    \State $idx \gets \arg\min(partitions\_sums)$
    \If{$(partitions\_sums[idx] + req.input\_len) + (1 + |partitions[idx]|) \times gen\_len > cache\_size$}
        \State $aborted\_requests \mathrel{+}= req$
    \Else
        \State $partitions[idx] \mathrel{+}= req$
        \State $partitions\_sums[idx] \mathrel{+}= req.input\_len$

        \If{$|partitions[idx]| == ubs$}
            \State $new\_batch \gets \Call{NewBatch}{partitions[idx]}$
            \State $micro\_batches \mathrel{+}= new\_batch$
            \State $partitions.pop(idx)$ 
            \State $partitions\_sums.pop(idx)$
        \EndIf
    \EndIf
\EndFor

\State \Return $micro\_batches$, $abort\_requests$
\end{algorithmic}
\end{algorithm}

\section{Further Discussion}
\subsection{MoE v.s. Dense Models}
The performance model and system optimizations proposed in this work are fully applicable to dense models. As discussed in \cref{sec:case_study}, MoE models present greater challenges with their higher memory-to-FLOPS ratio. This benefits them more from the system optimizations, which specifically aim to improve I/O efficiency and reduce pipeline bubbles. Dense models can benefit from these optimizations as well; however, they are more likely to be bottle-necked by CPU memory bandwidth during attention (depending on sequence length), where methods like sparse attention\cite{child2019generating, zhang2024h2o, tang2024quest} and quantized KV cache may offer more gains.

\subsection{Optimizer Overhead}
In \cref{sec:performance_model}, we introduced the optimization target \cref{eq:T} and the search space. For a given workload, model and hardware specification, the optimal policy can be generated offline through mixed integer linear programming (MILP), which takes less than a minute.

\section{Future Work}
\paragraph{Advanced performance model}
 \cm presented in this work is limited to hardware within a single node and does not account for GPU-GPU communication or multi-node communication, both of which are critical for more comprehensive distributed performance modeling. Additionally, with recent advances in leveraging KV cache sparsity for long-context inference~\cite{tang2024quest}, it becomes essential to incorporate these optimizations into the performance model. For example, when CPU attention emerges as the bottleneck, the KV cache budget can be adjusted to better balance CPU and GPU computation, enhancing overall system efficiency.

 \paragraph{Disk and other hardware support}
 \sys currently focuses on scenarios where GPU memory is limited but sufficient CPU memory is available to hold the model, highlighting the effectiveness of both \pipe and \cm. However, when CPU memory is insufficient to hold the entire model, disk offloading becomes essential. Moreover, supporting hardware such as TPUs and other accelerators is essential for extending the versatility of \sys to diverse computing environments.

\bibliographystyle{plain}
\bibliography{references}

\end{document}